\documentclass[10pt, conference, compsocconf]{IEEEtran}
\usepackage[pdftex]{graphicx}
\DeclareGraphicsExtensions{.png,.jpeg}
\usepackage[cmex10]{amsmath}
\usepackage{amsfonts}
\usepackage{amssymb}

\usepackage[cmex10]{amsmath}
\usepackage{array}
\usepackage{caption}
\usepackage{subfigure}
\begin{document}

%



\title{A Data-Driven Approach to Extract Connectivity Structures from Diffusion Tensor Imaging Data}




\author{\IEEEauthorblockN{Yu Jin\IEEEauthorrefmark{1}, Joseph F. JaJa\IEEEauthorrefmark{1}, Rong Chen\IEEEauthorrefmark{2}, Edward H. Herskovits\IEEEauthorrefmark{2}\\}
	\IEEEauthorblockN{\IEEEauthorrefmark{1}Institute for Advanced Computer Studies and Department of Electrical and Computer Engineering\\ University of Maryland, College Park, USA\\
	Email: yuj@umd.edu, joseph@umiacs.umd.edu \\
		\IEEEauthorrefmark{2}Department of Radiology, University of Maryland, Baltimore, Baltimore, USA\\
		Email: rong.chen.mail@gmail.com, ehh@ieee.org}
}


\maketitle

\begin{abstract}

Diffusion Tensor Imaging (DTI) is an effective tool for the analysis of structural brain connectivity in normal development and in a broad range of brain disorders. However efforts to derive inherent characteristics of structural brain networks have been hampered by the very high dimensionality of the data, relatively small sample sizes, and the lack of widely acceptable connectivity-based regions of interests (ROIs).  Typical approaches have focused either on regions defined by standard anatomical atlases that do not incorporate anatomical connectivity, or have been based on voxel-wise analysis, which results in loss of statistical power relative to structure-wise connectivity analysis.  In this work, we propose a novel, computationally efficient iterative clustering method to generate connectivity-based whole-brain parcellations that converge to a stable parcellation in a few iterations. Our algorithm is based on a sparse representation of the whole brain connectivity matrix, which reduces the number of edges from around a half billion to a few million while incorporating the necessary spatial constraints. We show that the resulting regions in a sense capture the inherent connectivity information present in the data, and are stable with respect to initialization and the randomization scheme within the algorithm. These parcellations provide consistent structural regions across the subjects of population samples that are homogeneous with respect to anatomic connectivity. Our method also derives connectivity structures that can be used to distinguish between population samples with known different structural connectivity. In particular, new results in structural differences for different population samples such as Females vs Males, Normal Controls vs Schizophrenia, and different age groups in Normal Controls are also shown.
\end{abstract}


\IEEEpeerreviewmaketitle

\section{Introduction}


Diffusion Magnetic Resonance  Imaging (MRI) technology  non-invasively reveals white matter fiber structures and provide a model of the brain fiber tracts at a relatively high resolution. This opens up new research opportunities to generate, explore and analyze complex brain networks derived from Diffusion Tensor Imaging (DTI) based structural connectivity information  \cite{bullmore2009complex,sporns2013structure,sporns2004organization}. Researchers have successfully applied graph theoretical analysis on specialized structural networks to shed light on differences between different population groups and on brain disorders such as dementia  \cite{dopper2014structural} and schizophrenia \cite{van2010aberrant}. Brain network analysis  requires a reasonably accurate anatomical segmentation of the cerebral cortex, called parcellation, in which structurally homogeneous regions constitute the nodes of the network. Traditional anatomical brain regions may not incorporate connectivity information, and are typically identified by the distribution of cell types \cite{amunts2007cytoarchitecture}, myelinated fibers  \cite{vogt1911myeloarchitektonik}, or neurotransmitter receptors \cite{zilles2009receptor}. Common widely-used anatomical brain parcellations include Brodman's areas \cite{brodmann1909vergleichende}, Automated Anatomical Labeling (AAL) \cite{tzourio2002automated}, and J\"{u}lich histological parcellations \cite{eickhoff2005new}. However, there are no generally accepted anatomical parcellations and atlases of the whole brain that are based purely on the anatomic brain connectivity information revealed by diffusion MRI data. Most of existing DTI-based parcellation studies focus on particular parts of the cerebral cortex, such as the human inferior parietal cortex complex (IPCC) \cite{ruschel2014connectivity}, the lateral parietal cortex \cite{mars2011diffusion}, the temporoparietal junction area (TPJ) \cite{mars2012connectivity}, the dorsal frontal cortex \cite{sallet2013organization}, the ventral frontal cortex \cite{neubert2014comparison}, cingulate and orbitofrontal cortex \cite{neubert2015connectivity}, and Broca's areas \cite{anwander2007connectivity}. The parcellations generated specifically for these regions have a small number of subregions but achieve high consistency among subjects of a population sample. Connectivity-based parcellation of the whole brain is challenging due to a number of factors that include: (i) the very large size of the connectivity matrix produced by tractography of each subject's DTI data; (ii) spatial constraints among the voxels of each region that must be respected in addition to the connectivity information; (iii) enforcing consistency for any structurally homogeneous population sample; and (iv) the lack of effective techniques to evaluate, and validate good parcellations. 
\par In this paper, we propose a novel iterative method based on spectral clustering applied to a sparse representation of the connectivity information which also incorporates the necessary spatial constraints. Our goal is to generate reproducible whole-brain parcellations based purely on DTI data, which are stable and subject-reproducible, achieve highly structurally homogeneous regions, and are consistent among structurally similar population samples.  Such parcellations can be used as the basis for conducting graph-theoretic analysis on the resulting anatomic connectivity networks. Our method uses probabilistic tractography to generate the connectivity matrix that represents connectivity strength between any two gray voxels. A sparse representation of the connectivity matrix is defined by a graph whose edges capture spatial connectivity within a small spatial neighborhood and whose edge weights provide a similarity measure of the connectivity profiles of the endpoints. We show that our method is effective in generating parcellations that are highly consistent among subjects in the same population sample and that capture anatomic connectivity patterns that can be used to distinguish between population samples with known structural differences. Moreover, the methods are computationally efficient and robust to various random factors. 
\par We note two particular works that are directly related to this paper. The first, reported by Craddock et al.  \cite{craddock2012whole}, focuses on a data-driven approach for generating atlases based on {\it resting-state functional} MRI. The main goal there is to parcellate the whole brain into coherent  regions of interests that are homogeneous in their resting-state functional connectivity (FC). They develop independently a graph formulation that is similar to ours, and apply spectral clustering in a straightforward way.  The resulting atlas, while better than several of the standard anatomical atlases in term of FC homogeneity, has similar characteristics to a random atlas. Moreover, the input size is significantly smaller than the size of the problem we are dealing with here. The second work reported in \cite{moreno2014hierarchical} addresses the same problem tackled in this paper and uses hierarchical clustering to generate a hierarchy of whole brain parcellations. Hierarchical clustering techniques have serious limitations since they use a local greedy strategy, and each successive refinement cannot modify the clustering determined in previous steps. In addition, the evaluation methodology carried out there is limited to either known results for small regions such as the inferior parietal cortex convexity or to other well-known cytoarchitectonic parcellations that do not incorporate the connectivity information provided by DTI. 
\par We summarize our main contributions in this paper as follows:
\begin{itemize}
	\item We develop efficient, scalable algorithms based on a sparse representation of the whole brain connectivity matrix, which reduces the number of edges from around a half billion to a few million while incorporating the necessary spatial constraints.
	\item For an arbitrary subject from a population sample and for any value $k$ of the number of regions, we show that our algorithm converges to a stable parcellation after a few iterations, defined by $k$ structurally homogeneous regions.
	\item Our parcellations of subjects within a population sample are consistent using any of a number of similarity metrics between parcellations of different subjects. 
	\item Our method captures  structural patterns to allow us to distinguish effectively between structurally different population groups such as Males vs Females, Normal Controls vs Schizophrenia, and different age groups in Normal Controls.
\end{itemize}
	\par The rest of the paper is organized as follows. We start in the next section by describing the data and tools used to generate the connectivity matrix of each subject. Our iterative method is described in Section III, while the stability and reproducibility results at the individual subject level or group level are covered in Section IV. Section V covers the discriminative power of the resulting parcellations. We end with a brief discussion in Section VI.

\begin{table}[t!]
	\caption{Subject Demographics}
	\label{Demographics}
	\centering
	\begin{tabular}{|c|c|c|m{3em}|m{3em}|m{3em}|c|}
		\hline
		\textbf{Subject Group} & \textbf{Male} & \textbf{Female} & \textbf{Age 18-30} & \textbf{Age 31-50} & \textbf{Age 51-60} & \textbf{Total}\\
		\hline
		Normal Controls & 41 & 35 & 23 & 28 & 25 & 76\\
		\hline
		Schizophrenia & 31 & 17 & 16 & 17 & 15 & 48\\
		\hline
	\end{tabular}
\end{table}

\section{Data and Preprocessing Steps}
\subsection{Data Acquisition}
Imaging was performed at the University of Maryland Center for Brain Imaging Research using a Siemens 3T TRIO MRI (Erlangen, Germany) system and 32 channel phase array head coil. The high-angular resolution diffusion imaging protocol was used to assess white matter integrity as measured by fractional anisotropy. Diffusion tensor data were collected using a single- shot, echo-planar, single refocusing spin-echo, T2-weighted sequence with a spatial resolution of 1.7$\times$1.7$\times$3.0mm. The sequence parameters were: TE/TR=87/8000ms, FOV=200mm, axial slice orientation with 50 slices and no gaps, 64 isotropically distributed diffusion weighted directions, two diffusion weighting values (b=0 and 700s/mm$^{2}$) and six b=0 images. These parameters were calculated using an optimization technique that maximizes the contrast to noise ratio for FA measurements. The total scan time was approximately 9 minutes per subject. For each subject, the image data consists of 70 volumes of 3D images of dimensions 128$\times$128$\times$53, each voxel representing 1.718mm$\times$1.718mm$\times$3mm brain volume. We collected data from 76 normal (NC) subjects and 48 schizophrenia (SZ) subjects. The subject demographics are shown in Table \ref{Demographics}.
\subsection{Nonlinear Registration}
The diffusion images of all subjects are registered to Montreal Neurological Institute (MNI) standard space using nonlinear registration package FNIRT in FSL \cite{andersson2007non}. The nonlinear registration process generates the warping coefficients that balance the similarity between the diffusion image and the standard MNI152 image, and the smoothness of the warping coefficients. The registration process facilitates group atlas generation and comparison with other standard atlases.
\subsection{Probabilistic tractography}
The preprocessing step of probabilistic tractography is used to model cross fiber distributions for each voxel through the  BEDPOSTX package in FSL \cite{behrens2007probabilistic}. Probabilistic tractography is processed through the diffusion toolbox in FSL \cite{behrens2003characterization}. The standard white matter atlas is specified as a seed region. The AAL mask is specified as the target region, which is the whole brain cortex region. We generate 50 streamlines from every voxel in a seed region. These streamlines are propagated following the cross fiber distribution computed from the preprocessing step. Curvature threshold is enforced to eliminate unqualified streamlines. The distance correction option is set to correct for the fact that the distribution drops as travel distance increases. The tractography output is a structural connectivity network modeled as a weighted graph where each node is a voxel in the target region space and each edge weight corresponds to relative connectivity strength in terms of the number of streamlines connecting the corresponding pair of voxels.

\begin{figure}[!t]
	\centering
	\includegraphics[width = .5\hsize]{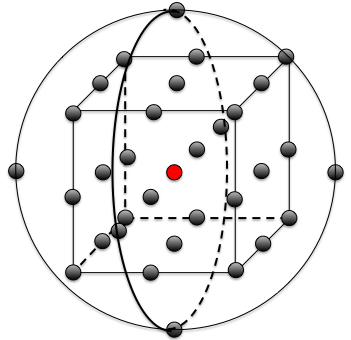}
	\caption{32 neighbors of voxel within sphere of radius $r = 2$.}
	\label{figure1}
\end{figure}
	
\section{Our Approach}
Our main method takes as input a subject's connectivity matrix. The number of voxels in the AAL mask is $155,794$ and the connectivity matrix is a sparse matrix of size $155,794 \times 155,794$. Given a positive integer value $k$, our problem is to parcellate the cerebral cortex into $k$ spatially contiguous regions, such that each region possesses a high degree of structural homogeneity. Moreover, these parcellations must be stable and reproducible, as well as, consistent among members of a population sample with similar connectivity patterns. We first introduce our notion of a \emph{connectivity profile} followed by a description of our method.
\subsection{Connectivity Profile}
For each voxel, the connectivity profile is the signature that discriminates a voxel from the rest of voxels based on connectivity. Parcellations are built by clustering voxels with similar connectivity profiles together. In principle, we can take the row of the connectivity matrix corresponding to a voxel as its connectivity profile, but that would be computationally expensive to process even if we compress each row into a list that contains only connectivity values above a certain threshold. In our approach, the connectivity profile of a voxel is computed as an array of weights, where each element represents the cumulative connectivity strengths of the voxel to a set of the regions determined initially by a predefined brain segmentation. Not only does the use of a coarser version of the connectivity profile leads to much more efficient computations, but it also helps to smooth out errors introduced by the tractography process through aggregation. More importantly, we will show later that our method converges to the same parcellation regardless of the initial segmentation used.
\par We explore several possibilities to initialize brain segmentations. An obvious choice is to use the regions of interests (ROIs) defined by any of the well-known anatomical atlases such as the 90 regions of the AAL-90 atlas. Note that the initial number of spatial regions is \emph{completely unrelated} to the number $k$ of parcellated regions and is used merely to initialize the connectivity profile of each voxel. Another possibility is to use a brain segmentation generated using a spatially constrained version of the k-means++ algorithm with  randomized centers \cite{arthur2007k}. The third possibility that we consider is to spatially segment the volume into almost equal-size sub-cubes. The last two segmentation methods result in any specified number of contiguous regions. We will show that our method results in consistent and similar parcellations regardless of which connectivity profile we use.
\subsection{Spatially Constrained Similarity Graph}
A spatial-constraint similarity graph, considerably sparser than the weighted graph defined by the connectivity matrix, is formed using spatial adjacency and the connectivity profiles as follows. The voxels define the nodes of our graph. Two nodes are connected by an edge if and only if the corresponding voxels lie within a sphere of radius $r$. In our implementation, we have used $r = 2$ such that the number of neighbors of any node is at most $32$ as shown in Fig. \ref{figure1}. Each edge is weighted by the similarity between the connectivity profiles of its end points. We can use any of several similarity metrics, including the correlation coefficient or the cosine function; our tests show that the results are very similar regardless of the similarity measure used.  We assume from now on that we are using the correlation coefficient as our similarity measure between the connectivity profiles of two voxels.
\begin{table}[t!]
	\centering
	\begin{tabular}{m{8cm}}
		\hline \hline
		\textbf{Algorithm1} Iterative Parcellation Method \\
		\hline	
		1. Generate the connectivity matrix of a subject using probabilistic tractography.\\
		2. Construct a spatial graph as a sparse representation of the 3-D brain.\\
		3.	Initialize a random spatially-coherent brain parcellation, to be used to define the connectivity profile of each voxel.\\
		\textbf{Repeat}\\
		4. Use the current brain parcellation to define the connectivity profiles of all the voxels based on the connectivity matrix.\\
		5.	Apply spectral clustering algorithm to generate the brain parcellation of a predefined level of granularity. \\
		6.	Measure the similarity between the new parcellation and the previous parcellation used to define connectivity profiles.\\
		\textbf{Until} the similarity measurement exceeds some threshold.\\
		7. Return the parcellation result.\\
		\hline \hline	
	\end{tabular}
\end{table}

\subsection{Minimum Graph-Cut Problem and Iterative Refinement}
Our parcellation algorithm starts by partitioning our spatial similarity graph into several subgraphs with the objective of minimizing  the total weight of the edges connecting the subgraphs subject to a constraint on the relative sizes of the subgraphs. More specifically, our objective function is to minimize the normalized cut rather than just the cut, which is standard in the literature (see for example \cite{von2007tutorial, ng2002spectral, yu2003multiclass}). This will more or less ensure that we won't have subgraphs with very few vertices. The subgraphs induce a spatial segmentation of the 3D image data, which is then used to redefine the connectivity profile of each voxel, after which we iterate until the generated parcellations are almost unchanged. Our algorithm results in a solution where the voxels within the same region have similar connectivity profiles and voxels across different regions are relatively dissimilar. The most efficient method to solve the graph cut problem during each iteration is spectral clustering  \cite{von2007tutorial, ng2002spectral, yu2003multiclass}. In particular, we use the normalized spectral clustering method, which can be summarized as follows, where $ W \in \mathbb{R}^{n\times n} $ is the weight matrix associated with the spatial similarity graph and $k$ is the number of desired regions. \par
\begin{itemize}
	\item Compute the normalized Laplacian matrix $L = D - W$. $D$ is the diagonal matrix with each element $D_{i,i} = \sum_{j=1 }^{n} W_{i,j}$.
	\item Compute the $k$ eigenvectors of $D^{-1/2}LD^{-1/2}$ corresponding to the smallest $k$ eigenvalues.
	\item Apply the k-means clustering algorithm on the rows of the eigenvectors to obtain the final clusters.
\end{itemize}
\par To make the clustering result consistent against the random initializations in the k-means step, we run the k-means++ 
algorithm \cite{arthur2007k} several times and choose the result with the minimum within-cluster sum of point-to-centroid distances \cite{matlabkmeans}. Note that each run of the k-means++ involves $155,794$ points (voxels) each of dimension $k$. \textbf{Algorithm1} provides a high-level description of our method.
\par By applying the spectral clustering algorithm, we expect voxels within the same region to possess successively higher degrees of similarity in terms of structural connectivity during successive iterations. This iterative refinement approach  converges to a stable parcellation as we will later show. At that point, we will also introduce a quantitative stopping criterion to be used to terminate the algorithm.
\section{Reproducibility and Stability Analysis}
This section presents the methodology used and the results achieved to illustrate the reproducibility of our results both at the individual subject level and at the group level. We start by introducing two well-known methods to quantitatively measure the similarity between two arbitrary clustering solutions of a dataset. 
\subsection{Parcellation Similarity Metrics}
We use the following metrics to measure the similarity between any two parcellations with the same level of granularity (that is, the same value of $k$). The cluster labels generated by our method are essentially arbitrary in the sense that regions with the same labels in two different parcellations are not necessarily spatially related. Moreover, as the level of granularity increases, we may not be able to determine a reasonable one-to-one mapping between the regions. In this paper, we will use the following two metrics.
\subsubsection{Normalized Mutual Information (NMI)}
Mutual information has been used in information theory to measure the relationship between any two probability distributions \cite{vinh2010information}. Essentially, it provides a measure of how similar the joint distribution of two random variables is to the product of their marginal distributions. The normalized mutual information (NMI) is an approximate discrete version commonly used to measure the similarity between pairs of clusters of a dataset. It has a value between $0$ and $1$, with the value $0$ indicating the two clusterings are completely independent of each other, whereas the value $1$ indicates that they are identical. The NMI between two parcellations $A$ and $B$ is defined as 
\begin{equation}
\label{NMI}
NMI\left( A, B\right) = \frac{MI\left( A, B\right) }{\left( H\left( A\right) + H\left( B\right) \right) / 2}
\end{equation}
\par The entropy for individual parcellations and the mutual information are approximated from the marginal and joint distributions as follows. $A_i$ is the set of voxels that are labeled as $i$ in parcellation $A$. Similarly, $B_j$ is the set of voxels that are labeled as $j$ in parcellation $B$.
\begin{equation}
\label{H_A}
H\left( A\right) = - \sum_{i=1}^{k}p\left( A_i\right) \log p\left( A_i\right)
\end{equation}
\begin{equation}
\label{H_B}
H\left( B\right) = - \sum_{j=1}^{k}p\left( B_j\right) \log p\left( B_j\right)
\end{equation}
\begin{equation}
\label{MI}
MI\left( A, B\right) = \sum_{i=1}^{k}\sum_{j=1}^{k} p\left( A_i, B_j\right) \log \left( \frac{p\left( A_i, B_j\right)}{p\left( A_i\right) p\left(B_j \right)}\right)
\end{equation}
\par The marginal probability for any label is approximated as the fraction of the number of voxels with that label over the total number of voxels. Similarly, the joint distribution $p\left( A_i, B_j\right)$ is computed as the fraction of the number of voxels with label $i$ in parcellation $A$ and with label $j$ in parcellation $B$ over the total number of voxels. Here the total number of voxels is the number of voxels in the AAL mask, which is the same for all parcellations.
\begin{equation}
\label{p_A}
p\left( A_i\right) = \frac{size\left( A_i\right) }{ \sum_{i = 1}^{k} size\left( A_i \right) }, p\left( B_j\right) = \frac{size\left( B_j\right) }{ \sum_{j = 1}^{k} size\left( B_i \right) }
\end{equation}

\begin{equation}
\label{p_AB}
p\left( A_i, B_j\right) = \frac{size\left( A_i \cap B_j\right) }{\sum_{i = 1}^{k} size\left( A_i \right)}
\end{equation}

\par As stated previously, if the parcellations are identical, except for label reordering, then the mutual information and the entropy for each parcellation are equal, and hence the resulting NMI is equal to $1$. The higher the value of the NMI, the more similar the two parcellations are.

\subsubsection{Dice's Coefficient}
Dice's coefficient measures the similarity directly from the clustering matrix $C \in \mathbb{R}^{n\times n} $ defined by 
\begin{equation}
C_{i,j} =\begin{cases}
1, & L_i = L_j \\
0, & L_i \neq L_j 
\end{cases}
\end{equation}
where $L$ is the vector that contains the label of every voxel. That is, the $\left( i,j\right) $ entry of the clustering matrix is equal to $1$ if, and only if, voxels $i$ and $j$ belong to the same region. Given the matrices corresponding to two parcellations, the Dice's coefficient is computed as twice the number of common nonzero entries normalized by the total number of nonzero entries in both clustering matrices  \cite{dice1945measures}.  Dice's coefficient is always between $0$ and $1$, and the larger it is, the more similar the two parcellations are.\\
\par Both NMI and Dice's coefficient capture the similarity between two parcellations of any level of granularity. But for NMI, the joint distribution $p\left( A_i,B_j \right) $ is in general greater than the product of the marginal distributions $p\left( A_i \right) p\left( B_j \right) $, which may cause NMI to overestimate the similarity between the parcellations. We note that in general NMI is larger than the Dice's coefficient.

\subsection{Stability and Reproducibility of Subject Parecellations}
The main factors that affect the parcellations generated by our algorithm are the choice of the brain segmentation that is 
used to define connectivity profiles and the random initialization of the k-means++ algorithm used in the last step of spectral clustering. The effect of random initialization could be mitigated by running the k-means++ initialization \cite{arthur2007k} several times, as stated before. Here we consider only the effect of the initial brain segmentation used to define the connectivity profiles. Note that the connectivity profiles encapsulate the only information we have from the DTI data for each subject since the rest of the information captured by the spatial similarity graph does not involve anything related to the connectivity data.  
\par The initial brain segmentation can be defined as any arbitrary spatial segmentation of the brain mask. However it would be more intuitive to use initial segmentations with comparable region sizes. Note that the number of regions in the 
initial segmentation is completely independent of the desired number $k$ of parcellated regions.  The main result of this section is that, regardless of the initial segmentation and for any value of $k$, our algorithm will converge to a stable parcellation for each subject that captures the critical connectivity information embodied in the DTI data.

\par
\begin{figure}
	\centering
	\includegraphics[width = 0.9\hsize]{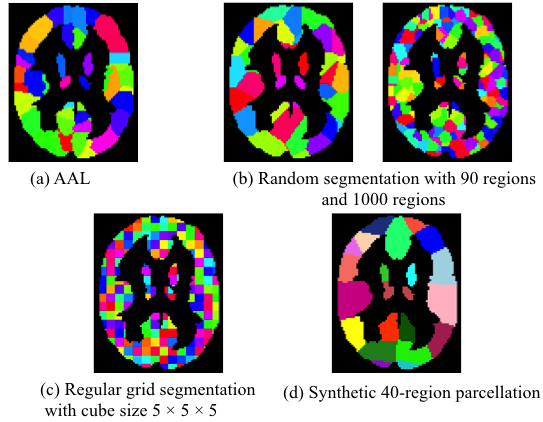}
	\caption{Brain segmentations used to define connectivity profiles.}
	\label{figure2}
\end{figure}

 The following brain segmentations, shown in Fig. \ref{figure2}, are used to define initial connectivity profiles that are used to generate 40-region parcellations.

\subsubsection{Automated Anatomical Labeling (AAL)}
The AAL atlas defines 90 anatomical regions with 45 volumes of interest in each hemisphere, which were delineated following the courses of the main sulci of the brain. In fact, we have used the AAL mask to define cerebral cortex to be parcellated. Here we use it as the initial segmentation that defines the connectivity profiles of all voxels.  
\subsubsection{Random Spatial Segmentation}
We generate a random spatial segmentation with any level of granularity using the k-means++ algorithm, based only on spatial coordinate. The purpose is to generate segmentations that have regions that are spatially contiguous and compact. Random initialization of the k-means++ using 90, 1000, and 2000 regions were generated.
\subsubsection{Regular Grid Segmentation}
A regular grid segmentation consists of a set of almost equal-sized cubes that cover the whole brain. The cube size determines the granularity of the segmentation. We set the cube size to 5 and therefore, this segmentation consists of 1,987 cubes that cover all brain voxels.
\subsubsection{Synthetic Parcellations}
The synthetic parcellations are generated from the similarity graph in which the weights of all the edges are set to 
1. A similar approach was reported in \cite{craddock2012whole}, which concludes that the synthetic parcellations are almost as good as real parcellations in terms of FC cluster homogeneity. We use the synthetic parcellation with the same number of regions to define the connectivity profile and show that starting from the same synthetic parcellation, our iterative method will incorporate the underlying connectivity information and converge to the subject's characteristic parcellations. 
\par For each subject, we show that our algorithm will yield essentially the same parcellation for all these initial segementations. The NMI and Dice's coefficient are computed between all pairs of parcellations generated from different brain segmentations after each iteration. Tables \ref{NMI_1} through \ref{Dice_4} show the corresponding results. 
\begin{table}[t!]
	\centering
	\begin{tabular}{m{8cm}}
		\hline \hline
		\textbf{Abbreviations}\\
		\hline
		AAL: Automatic Anatomical Labeling. \\
		R\#: Random brain segmentation with \# number of regions.\\
		Grid: Regular grid segmentaton with grid size $5 \times 5 \times 5$. \\
		S\#: Synthetic parcellation generated from spatial-constrained similarity graph with all edges' weights as $1$.\\
		\hline \hline
	\end{tabular}
\end{table}

\begin{table}[t!]
	\caption{NMI Between Parcellations After $1^{st}$ Iteration }
	\label{NMI_1}
	\centering
	\begin{tabular}{|m{3em}|c|c|c|c|c|c|}
		\hline
		\textbf{Segmen-tation} & AAL    & R90    & R1000  & R2000  & Grid   & S40    \\ \hline
		AAL          & 1.0000 & 0.8673 & 0.8791 & 0.8497 & 0.8734 & 0.8568 \\ \hline
		R90          & 0.8673 & 1.0000 & 0.9009 & 0.8622 & 0.8849 & 0.8804 \\ \hline
		R1000        & 0.8791 & 0.9009 & 1.0000 & 0.8774 & 0.9039 & 0.8657 \\ \hline
		R2000        & 0.8497 & 0.8622 & 0.8774 & 1.0000 & 0.8855 & 0.8433 \\ \hline
		Grid         & 0.8734 & 0.8849 & 0.9039 & 0.8855 & 1.0000 & 0.8666 \\ \hline
		S40          & 0.8568 & 0.8804 & 0.8657 & 0.8433 & 0.8666 & 1.0000 \\ \hline
	\end{tabular}
\end{table}

\begin{table}[t!]
	\caption{NMI Between Parcellations After $2^{nd}$ Iteration }
	\label{NMI_2}
	\centering
	\begin{tabular}{|m{3em}|c|c|c|c|c|c|}
		\hline
		\textbf{Segmen-tation} & AAL    & R90    & R1000  & R2000  & Grid   & S40    \\ \hline
		AAL   & 1.0000 & 0.9173 & 0.9085 & 0.9117 & 0.8999 & 0.8990 \\ \hline
		R90   & 0.9173 & 1.0000 & 0.9042 & 0.9031 & 0.8880 & 0.8954 \\ \hline
		R1000 & 0.9085 & 0.9042 & 1.0000 & 0.9018 & 0.8971 & 0.8908 \\ \hline
		R2000 & 0.9117 & 0.9031 & 0.9018 & 1.0000 & 0.8840 & 0.8780 \\ \hline
		Grid  & 0.8999 & 0.8880 & 0.8971 & 0.8840 & 1.0000 & 0.8980 \\ \hline
		S40   & 0.8990 & 0.8954 & 0.8908 & 0.8780 & 0.8980 & 1.0000 \\ \hline
	\end{tabular}
\end{table}

\begin{table}[t!]
	\caption{NMI Between Parcellations After $3^{rd}$ Iteration }
	\label{NMI_3}
	\centering
	\begin{tabular}{|m{3em}|c|c|c|c|c|c|}
		\hline
		\textbf{Segmen-tation} & AAL    & R90    & R1000  & R2000  & Grid   & S40    \\ \hline
		AAL   & 1.0000 & 0.9194 & 0.8940 & 0.9412 & 0.8985 & 0.9035 \\ \hline
		R90   & 0.9194 & 1.0000 & 0.9050 & 0.9266 & 0.9242 & 0.9266 \\ \hline
		R1000 & 0.8940 & 0.9050 & 1.0000 & 0.9103 & 0.8899 & 0.8992 \\ \hline
		R2000 & 0.9412 & 0.9266 & 0.9103 & 1.0000 & 0.9087 & 0.9064 \\ \hline
		Grid  & 0.8985 & 0.9242 & 0.8899 & 0.9087 & 1.0000 & 0.9101 \\ \hline
		S40   & 0.9035 & 0.9266 & 0.8992 & 0.9064 & 0.9101 & 1.0000 \\ \hline
	\end{tabular}
\end{table}

\begin{table}[t!]
	\caption{NMI Between Parcellations After $4^{th}$ Iteration }
	\label{NMI_4}
	\centering
	\begin{tabular}{|m{3em}|c|c|c|c|c|c|}
		\hline
		\textbf{Segmen-tation} & AAL    & R90    & R1000  & R2000  & Grid   & S40    \\ \hline
		AAL   & 1.0000 & 0.9405 & 0.9052 & 0.9382 & 0.9004 & 0.9258 \\ \hline
		R90   & 0.9405 & 1.0000 & 0.9181 & 0.9211 & 0.9141 & 0.9494 \\ \hline
		R1000 & 0.9052 & 0.9181 & 1.0000 & 0.8949 & 0.8828 & 0.9148 \\ \hline
		R2000 & 0.9382 & 0.9211 & 0.8949 & 1.0000 & 0.9225 & 0.9075 \\ \hline
		Grid  & 0.9004 & 0.9141 & 0.8828 & 0.9225 & 1.0000 & 0.9020 \\ \hline
		S40   & 0.9258 & 0.9494 & 0.9148 & 0.9075 & 0.9020 & 1.0000 \\ \hline
	\end{tabular}
\end{table}

\par The above tables show that similarity, in terms of NMI or Dice's coefficients, between all pairs of parcellations from different brain segmentations increase with the number of iteration. After the $4^{th}$ iteration, most of NMI values are  
\begin{table}[ht!]
	\caption{Dice's Coefficient Between Parcellations After $1^{st}$ Iteration }
	\label{Dice_1}
	\centering
	\begin{tabular}{|m{3em}|c|c|c|c|c|c|}
		\hline
		\textbf{Segmen-tation} & AAL    & R90    & R1000  & R2000  & Grid   & S40    \\ \hline
		AAL   & 1.0000 & 0.7448 & 0.7709 & 0.7083 & 0.7666 & 0.7230 \\ \hline
		R90   & 0.7448 & 1.0000 & 0.8374 & 0.7456 & 0.8046 & 0.7810 \\ \hline
		R1000 & 0.7709 & 0.8374 & 1.0000 & 0.7778 & 0.8396 & 0.7553 \\ \hline
		R2000 & 0.7083 & 0.7456 & 0.7778 & 1.0000 & 0.7874 & 0.7010 \\ \hline
		Grid  & 0.7666 & 0.8046 & 0.8396 & 0.7874 & 1.0000 & 0.7585 \\ \hline
		S40   & 0.7230 & 0.7810 & 0.7553 & 0.7010 & 0.7585 & 1.0000 \\ \hline
	\end{tabular}
\end{table}

\begin{table}[ht!]
	\caption{Dice's Coefficient Between Parcellations After $2^{nd}$ Iteration }
	\label{Dice_2}
	\centering
	\begin{tabular}{|m{3em}|c|c|c|c|c|c|}
		\hline
		\textbf{Segmen-tation} & AAL    & R90    & R1000  & R2000  & Grid   & S40    \\ \hline
		AAL   & 1.0000 & 0.8557 & 0.8378 & 0.8483 & 0.8142 & 0.8236 \\ \hline
		R90   & 0.8557 & 1.0000 & 0.8203 & 0.8199 & 0.7852 & 0.8068 \\ \hline
		R1000 & 0.8378 & 0.8203 & 1.0000 & 0.8205 & 0.8097 & 0.7936 \\ \hline
		R2000 & 0.8483 & 0.8199 & 0.8205 & 1.0000 & 0.7737 & 0.7717 \\ \hline
		Grid  & 0.8142 & 0.7852 & 0.8097 & 0.7737 & 1.0000 & 0.8099 \\ \hline
		S40   & 0.8236 & 0.8068 & 0.7936 & 0.7717 & 0.8099 & 1.0000 \\ \hline
	\end{tabular}
\end{table}

\begin{table}[ht!]
	\caption{Dice's Coefficient Between Parcellations After $3^{rd}$ Iteration }
	\label{Dice_3}
	\centering
	\begin{tabular}{|m{3em}|c|c|c|c|c|c|}
		\hline
		\textbf{Segmen-tation} & AAL    & R90    & R1000  & R2000  & Grid   & S40    \\ \hline	
		AAL   & 1.0000 & 0.8543 & 0.8007 & 0.9021 & 0.8047 & 0.8197 \\ \hline
		R90   & 0.8543 & 1.0000 & 0.8316 & 0.8759 & 0.8700 & 0.8787 \\ \hline
		R1000 & 0.8007 & 0.8316 & 1.0000 & 0.8352 & 0.7932 & 0.8103 \\ \hline
		R2000 & 0.9021 & 0.8759 & 0.8352 & 1.0000 & 0.8317 & 0.8341 \\ \hline
		Grid  & 0.8047 & 0.8700 & 0.7932 & 0.8317 & 1.0000 & 0.8378 \\ \hline
		S40   & 0.8197 & 0.8787 & 0.8103 & 0.8341 & 0.8378 & 1.0000 \\ \hline
	\end{tabular}
\end{table}

\begin{figure}
	\centering
	\includegraphics[width = 0.7\hsize]{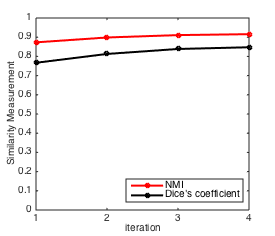}
	\caption{Subject reproducibility after each iteration.}
	\label{figure3}
\end{figure}

\begin{table}[ht!]
	\caption{Dice's Coefficient Between Parcellations After $4^{th}$ Iteration }
	\label{Dice_4}
	\centering
	\begin{tabular}{|m{3em}|c|c|c|c|c|c|}
		\hline
		\textbf{Segmen-tation} & AAL    & R90    & R1000  & R2000  & Grid   & S40    \\ \hline
		AAL   & 1.0000 & 0.9070 & 0.8279 & 0.8936 & 0.8125 & 0.8786 \\ \hline
		R90   & 0.9070 & 1.0000 & 0.8526 & 0.8547 & 0.8387 & 0.9247 \\ \hline
		R1000 & 0.8279 & 0.8526 & 1.0000 & 0.7960 & 0.7653 & 0.8472 \\ \hline
		R2000 & 0.8936 & 0.8547 & 0.7960 & 1.0000 & 0.8699 & 0.8290 \\ \hline
		Grid  & 0.8125 & 0.8387 & 0.7653 & 0.8699 & 1.0000 & 0.8171 \\ \hline
		S40   & 0.8786 & 0.9247 & 0.8472 & 0.8290 & 0.8171 & 1.0000 \\ \hline
	\end{tabular}
\end{table} 
\noindent above $0.90$ and most of Dice's coefficients are above $0.80$, which indicates very consistent parcellations. The iterative method mitigates the random effect caused by the initial arbitrary segmentations and leads to stable parcellations regardless of the initial definition of connectivity profiles. Note that the k-means++ step of our algorithm introduces a small uncertainty, which explains the few deviations in the tables above. However, it is clear that the parcellations generated at the end of third and fourth iterations are very close to each other. Fig. \ref{figure3} illustrates the increase of  the average, over all the different intial segmentations, of NMI and Dice's coefficient  after each iteration.
\par Table \ref{NMI_consecutive} and \ref{Dice_consecutive} show the similarity between parcellations in consecutive iteration stages for a given initial segmentation. Taking into consideration the uncertainty introduced by the k-means++ step of our algorithm, it is clear that successive iterations of the algorithm generate more similar parcellations, for any of the initialization methods of the connectivity profiles.

\begin{table}[t!]
	\caption{NMI Between Parcellations in Consecutive Iteration Stages }
	\label{NMI_consecutive}
	\centering
	\begin{tabular}{|c|c|c|c|}
		\hline
		\textbf{Segmentation} & \textbf{$1^{st}$ / $2^{nd}$} & \textbf{$2^{nd}$ / $3^{rd}$} & \textbf{$3^{rd}$ / $4^{th}$} \\ \hline
		AAL          & 0.9131    & 0.9353    & 0.9539    \\ \hline
		R90          & 0.9125    & 0.9325    & 0.9631    \\ \hline
		R1000        & 0.9199    & 0.9292    & 0.9198    \\ \hline
		R2000        & 0.8873    & 0.9224    & 0.9314    \\ \hline
		Grid         & 0.9185    & 0.9380    & 0.9267    \\ \hline
		S40          & 0.9151    & 0.9341    & 0.9486    \\ \hline
	
	\end{tabular}
\end{table}

\begin{table}[t!]
	\caption{Dice's coefficient Between Parcellations in Consecutive Iteration Stages }
	\label{Dice_consecutive}
	\centering
	\begin{tabular}{|c|c|c|c|}
		\hline
		\textbf{Segmentation} & \textbf{$1^{st}$ / $2^{nd}$} & \textbf{$2^{nd}$ / $3^{rd}$} & \textbf{$3^{rd}$ / $4^{th}$} \\ \hline
		AAL   & 0.8448 & 0.8886 & 0.9179 \\ \hline
		R90   & 0.8402 & 0.8838 & 0.9475 \\ \hline
		R1000 & 0.8714 & 0.8787 & 0.8495 \\ \hline
		R2000 & 0.7997 & 0.8556 & 0.8697 \\ \hline
		Grid  & 0.8700 & 0.8960 & 0.8709 \\ \hline
		S40   & 0.8520 & 0.8813 & 0.9124 \\ \hline
	\end{tabular}
\end{table}

\subsection{Group Consistency and Atlas Generation}
Table \ref{Diff_healthy} shows the average similarity between every pair of parcellations from subjects in the NC group. As can be seen from entries in this table, the parcellations are reasonably consistent within the NC group; similar results hold for the SZ group.
\par Table \ref{Diff_healthy_random} shows the average similarity between a random parcellation and the parcellations generated for the subjects in the NC group. As can be seen from the column of the Dice coefficients, our generated parcellations are significantly different from random parcellations. As mentioned before, the NMI coefficients tend to overestimate the similarity between the parcellations, and hence the slightly higher numbers in the second column of Table \ref{Diff_healthy_random}, but still significantly lower than the similarity of the generated parcellations between the subjects of the NC group (Table \ref{Diff_healthy}). 

\begin{table}[t!]
	\caption{Average Similarity Between Parcellations of Different Subjects within the NC Group}
	\label{Diff_healthy}
	\centering
	\begin{tabular}{|c|c|c|}
		\hline
		\textbf{Number of regions} & \textbf{\hspace{1cm}NMI\hspace{1cm}} & \textbf{Dice's Coefficient} \\ \hline
		40  & 0.7734 & 0.5503 \\ \hline
		50  & 0.7786 & 0.5323 \\ \hline
		60  & 0.7939 & 0.5507 \\ \hline
		70  & 0.7988 & 0.5415 \\ \hline
		90  & 0.8040 & 0.5326 \\ \hline
		120 & 0.8151 & 0.5287 \\ \hline
	\end{tabular}
\end{table}

\begin{table}[t!]
	\caption{ Average Similarity Between Parcellations of Subjects within the NC Group and Randomly Generated Parcellation}
	\label{Diff_healthy_random}
	\centering
	\begin{tabular}{|c|c|c|}
		\hline
		\textbf{Number of regions} & \textbf{\hspace{1cm}NMI\hspace{1cm}} & \textbf{Dice's Coefficient} \\ \hline
		40  & 0.6923 & 0.3994 \\ \hline
		50  & 0.6857 & 0.3679 \\ \hline
		60  & 0.7140 & 0.3871 \\ \hline
		70  & 0.7164 & 0.3771 \\ \hline
		90  & 0.7393 & 0.3995 \\ \hline
		120 & 0.7452 & 0.3720 \\ \hline
	\end{tabular}
\end{table}

\par \textit{Atlas generation}: We employ the following atlas generation procedure to further validate within-group consistency. In generating our parcellations, regions are labeled randomly; therefore, regions with the same index are not necessarily spatially matched. The first step of atlas generation is to align all parcellations to a reference parcellation that is randomly chosen from the group. We relabel each of the regions using the region index of the reference parcellation that shares the largest overlapped area. For a group of $N$ relabeled subjects, we generate an atlas as follows. For each voxel, we associate a vector of length $N$ consisting of the label index from each subject. We set the voxel's label to be the most frequent index in its vector, thereby generating an atlas as shown in Fig. 4(a).
\begin{figure}[t!]
	\centering
	\includegraphics[width = 0.8\hsize]{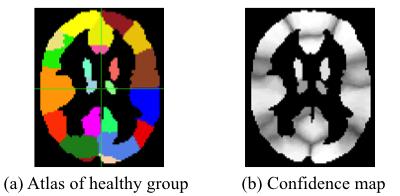}
	\caption{Atlas and confidence map for the NC group.}
	\label{figure4}
\end{figure}
\par The confidence map is a gray-scale image, where the gray level of each voxel represents the uncertainty of the labeling across all subjects, in terms of the proportion of the frequent index in the $N$-length vector. The confidence map in Fig. 4(b) shows that for almost all voxels, except possibly along the region boundaries, most subjects are consistently labeled as indicated by the atlas.

\section{Discriminative Analysis}
In this section, we show how our parcellations can be used to shed light on structural differences between different experimental groups. We have selected cases that were known to have significant differences in white matter integrity and structural networks. We include a discussion of three significant different groups: Male vs Female, Age groups, and SZ vs NC. The subject demographics in our data are shown in Table \ref{Demographics}.
\begin{table}[t!]
	\caption{ P-value and T-statistic for Gender Study within the NC Group}
	\label{p_gender}
	\centering
	\begin{tabular}{|c|c|c|}
		\hline
		\textbf{Similarity comparison} & \textbf{\hspace{0.5cm}p-value\hspace{0.5cm}} & \textbf{t-statistic} \\ \hline
		Male vs Male-female   & 9.0286e-5  & -3.9205 \\ \hline
		Female vs Male-female & 1.4025e-8  & 5.6911  \\ \hline
		Male vs Female        & 3.3752e-20 & -9.2766 \\ \hline
	\end{tabular}
\end{table}

\begin{table}[t!]
	\caption{ P-value and T-statistic for Age Study within the NC Group}
	\label{p_age}
	\centering
	\begin{tabular}{|c|c|c|}
		\hline
		\textbf{Similarity comparison} & \textbf{\hspace{0.5cm}p-value\hspace{0.5cm}} & \textbf{t-statistic} \\ \hline
		Group I vs Group I-II     & 0.5133    & -0.6539 \\ \hline
		Group II vs Group I-II    & 0.5566    & 0.5880  \\ \hline
		Group I vs Group II 	& 0.2009    & -1.2798  \\ \hline
		Group II vs Group II-III  & 2.4175e-4 & 3.6800  \\ \hline
		Group III vs Group II-III & 0.0028    & -2.9921 \\ \hline
		Group II vs Group III & 6.6455e-11    & 6.5814\\ \hline
	\end{tabular}
\end{table}

\begin{table}[t!]
	\caption{ P-value and T-statistic for Schizophrenic Study}
	\label{p_patient}
	\centering
	\begin{tabular}{|c|c|c|}
		\hline
		\textbf{Similarity comparison} & \textbf{\hspace{0.5cm}p-value\hspace{0.5cm}} & \textbf{t-statistic} \\ \hline
		NC vs NC-SZ & 2.9995e-89  & 20.2514  \\ \hline
		SZ vs NC-SZ      & 1.4025e-8   & -10.4867 \\ \hline
		NC vs SZ         & 1.1636e-198 & 30.9687  \\ \hline
	\end{tabular}
\end{table}

\par We adopt two strategies to discriminate among experimental groups. The first strategy focuses on the heterogeneity of the parcellations within a group sample and is based on the pair-wise similarities between all pairs of parcellations in a group. As shown in the previous section, our parcellations are consistently labeled across subjects of a population sample except for some boundary voxels. The boundary differences reflected by the pair-wise similarity may be used to determine some features that are specific to particular subgroups. In particular, we will show that the parcellations of the subjects in the SZ group have substantially more variability that those of the NC group and that healthy males seem to exhibit more heterogeneity within their group than healthy females do.

\par  The other strategy is to analyze the structural connectivity network built from the parcellations and tractography results, where the nodes correspond to the parcellation regions and the  edge weights correspond to the cumulative connectivity strength between voxels in the two regions; this strategy is commonly used in the literature  \cite{ingalhalikar2014sex, gong2009age}. Our iterative method generates parcellations where voxels within the same region share similar connectivity profiles that are defined as the accumulated connectivity strength to every other region. Hence the parcellations obtained are consistent with the structural connectivity network where the connectivity pattern of each node summarizes the connectivity profile of the voxels in that region. The ``connectome" analysis shows more powerful discriminative ability of our parcellations than using existing anatomical atlases. 
\subsection{Similarity-based Analysis}
The analysis is based solely on pair-wise similarity between pairs of parcellations. We start by analyzing the similarities relative to female and male subgroups of the NC sample using parcellations with 90 regions and NMI as the similarity measure. Results corresponding to other values of $k$ or to the use of Dice's coefficient as a similarity measure exhibit the same patterns.
\par A two-sample t-test was performed on the pair-wise similarity between parcellations within the female subgroup, the male subgroup, and pair-wise similarity between parcellations from different groups. The p-value and t-statistics are shown in Table \ref{p_gender}.
\par Our results indicate that the similarity of parcellations of either healthy females or healthy males is significantly different that the similarity between a female parcellation and a male parcellation. More importantly, the last row of Table \ref{p_gender} indicates that the female parcellations are much closer to each other that the male parcellations, which may indicate more structural brain heterogeneity among the male subjects than among the female subjects. 
\par In the age study, we divide the NC sample into three age groups, which are: Group I: Age 18-29, 23 subjects; Group II: Age 30-49, 28 subjects; Group III: Age 50-62, 25 subjects. The p-value and t-statistics are shown in table \ref{p_age}. For parcellations in Group I and II, their similarities did not show significant differences. But parcellation similarities within Group II have significant differences than the similarity between parcellations in Group II and Group III. And parcellations in Group III are more heterogeneous than those in Group I and Group II. 
\par Perhaps more interesting is the similarity comparison of the parcellations of NC vs SZ groups, illustrated in  \ref{p_patient}. These results clearly show that the parcellations within the SZ group show much more heterogeneity than those for the NC group. 

\subsection{Connectome Analysis}
\begin{figure}[t!]
	\centering
	\includegraphics[width = 0.8\hsize]{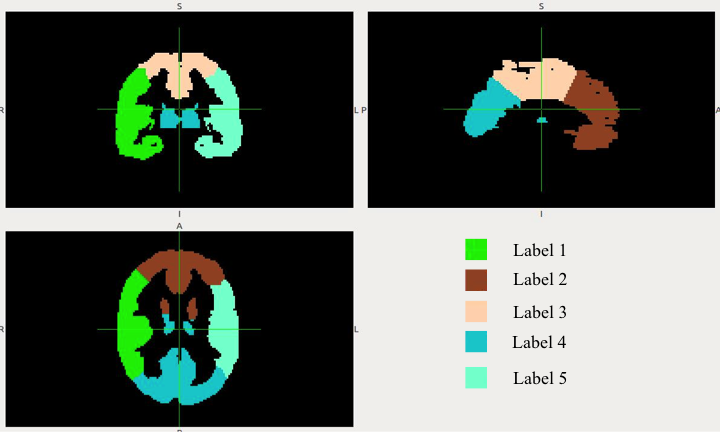}
	\caption{Parcellation with 5 regions.}
	\label{figure5}
\end{figure}
There is much evidence supporting that schizophrenia is a disorder related to brain connectivity. Our previous work analyzed the structural connectivity network based on individual parcellations refined from the AAL atlas to discriminate schizophrenic and normal control groups with high accuracy  \cite{qi2015}. We apply the same strategy to discriminate among the two groups using the 5-region parcellations generated from our iterative approach as shown in Fig. \ref{figure5}. The reason we choose a small number of regions is the high consistency across subjects, and because regions can be trivially mapped spatially, one-to-one, between any pair of parcellations.

\par We first relabel all parcellations based on a randomly selected subject. The connectomes are built by defining nodes as regions in the parcellation and edge weights represent cumulative connectivity strength between regions. Table \ref{p_conn} shows the p-value and t-statistcs of pair-wise connectivity between the two groups.
\par A large portion of pair-wise connectivity shows significant differences between the two groups. Moreover, most pair-wise  connectivity strengths of NC subjects are greater than those of SZ subjects, a fact that is consistent with the previous findings that SZ subjects have decreased inter-hemispheric and intra-hemispheric connectivity \cite{herskovits2015edge}. We select the three pairs with the most significant p-values and use their connectivity values as features to train a support vector machine classifier. We test our classifier using a 10-fold cross-validation and are able to achieve up to $75\%$ accuracy, which is significantly better than our earlier result in \cite{qi2015}. 

\begin{table}[t!]
	\caption{ P-value and T-statistic of Pair-wise Connectivity Between Normal Controls and Schizophrenic Groups}
	\label{p_conn}
	\centering
	\begin{tabular}{|c|c|c|c|c|c|}
		\hline
		\textbf{Label} & 1      & 2      & 3       & 4       & 5      \\ \hline
		1     & 0.0137 & 0.0395 & 0.0038  & 0.0989  & 0.4954 \\ \hline
		2     & 0.0300 & 0.0279 & 0.0029  & 0.0514  & 0.0058 \\ \hline
		3     & 0.0029 & 0.0042 & 4.11e-4 & 3.04e-5 & 0.0019 \\ \hline
		4     & 0.0798 & 0.0579 & 3.14e-5 & 0.0775  & 0.1127 \\ \hline
		5     & 0.4928 & 0.0083 & 0.0028  & 0.1216  & 0.2377 \\ \hline \hline
	\end{tabular}
	
	\begin{tabular}{|c|c|c|c|c|c|}
		\hline
		\textbf{Label} & 1      & 2      & 3       & 4       & 5      \\ \hline
		1 & 2.5019  & 2.0809 & 2.9479 & 1.6630 & -0.6839 \\ \hline
		2 & 2.1963  & 2.2250 & 3.0391 & 1.9671 & 2.8055  \\ \hline
		3 & 3.0437  & 2.9197 & 3.6325 & 4.3330 & 3.1707  \\ \hline
		4 & 1.7666  & 1.9149 & 4.3244 & 1.7806 & 1.5976  \\ \hline
		5 & -0.6879 & 2.6846 & 3.0493 & 1.5588 & 1.1865  \\ \hline
	\end{tabular}
\end{table}
\par
\begin{figure}[t!]
	\centering
	\includegraphics[width = 0.8\hsize]{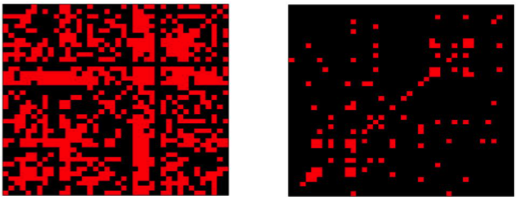}
	\caption{Binary maps where entries in red color have p-values  $<$0.05 and $<$0.00005 respectively in terms of connectivity strengths between the two population groups using our 40-region parcellations. }
	\label{figure6}
\end{figure}

\begin{figure}[t!]
	\centering
	\includegraphics[width = 0.8\hsize]{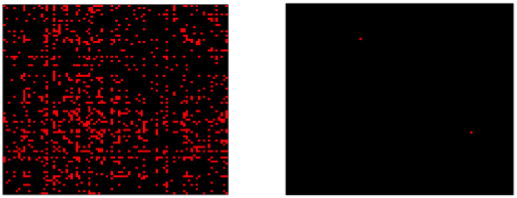}
	\caption{Binary map where entries in red color have p-values  $<$0.05 and $<$0.00005 respectively between connectivity strengths of the two population groups using the AAL atlas.}
	\label{figure7}
\end{figure}

 We also carried out an additional test to confirm the discriminative capabilities of our parcellations. Consider 40-region parcellations for the two population samples and the corresponding structural connectivity networks. For each edge, we perform a two-sample t-test between the sequence of connectivity strengths of the NC group and that of the SZ group. We find that many of the edges result in p-values less than 0.00005 as shown in Fig. \ref{figure6}. Fig. \ref{figure7} shows the corresponding results when we use the AAL atlas and determine connectivity strengths on the corresponding edges (pairs of regions). As shown by the binary maps, the proportion of entries in the AAL-based network which have significant connectivity strength difference between healthy controls and schizophrenic subjects is much smaller than that those obtained through the network built from our 40-region parcellations. 
\par It seems clear that our pacellations seem to effectively capture the inherent connectivity information present in the DTI data and hence are more suitable for studying structural connectivity than anatomical atlases.


\section{Conclusion}
We herein propose a sparse representation of the connectivity information derived from DTI data and a novel method that generates whole-brain parcellations for any number $k$ of regions. Our method is computationally efficient and is able to consistently generate stable and reproducible parcellations that seem to capture inherent structural patterns present in the data. The results are validated through the use of a number of methods, including subject reproducibility, group consistency, and discriminative characteristics between different population groups.

\section*{Acknowledgment}
We gratefully acknowledge funding provided by The University of Maryland/Mpowering the State through the Center for Health-related Informatics and Bioimaging (CHIB). 

\newcommand{\BIBdecl}{\setlength{\itemsep}{0.00 em}}
\bibliographystyle{IEEEtran}
\bibliography{bigdata}


\end{document}